\documentclass[onecolumn,showpacs]{revtex4}

\usepackage{graphicx}
\usepackage{dcolumn}
\usepackage{bm}

\input epsf

\begin{document}

\title{\Large Gravitational Collapse in Higher Dimensional Husain Space-Time}

\author{\bf Ujjal
Debnath$^1$\footnote{ujjaldebnath@yahoo.com},~ Narayan Chandra Chakraborty$^2$ and Subenoy
Chakraborty$^2$\footnote{subenoyc@yahoo.co.in}}

\affiliation{$^1$Department of Mathematics, Bengal
Engineering and Science University, Shibpur, Howrah-711 103,
India\\ $^2$Department of Mathematics, Jadavpur University,
Calcutta-32, India.\\ }

\date{\today}

\begin{abstract}
We investigate exact solution in higher dimensional Husain model
for a null fluid source with pressure $p$ and density $\rho$ are
related by the following relations (i) $p=k\rho$, (ii)
$p=k\rho-\frac{B(v)}{\rho^{\alpha}}$ (variable modified
Chaplygin) and (iii) $p=k\rho^{\alpha}$ (polytropic). We have
studied the nature of singularity in gravitational collapse for
the above equations of state and also for different choices of the
of the parameters $k$ and $B$ namely, (i) $k=0$, $B=$ constant
(generalized Chaplygin), (ii) $B=$ constant (modified Chaplygin).
It is found that the nature of singularity is independent of
these choices of different equation of state except for variable
Chaplygin model. Choices of various parameters are shown in
tabular form. Finally, matching of Szekeres model with exterior
Husain space-time is done.
\end{abstract}

\pacs{04.20~Jb,~~04.20~Dw,~~97.60. Lf}

\maketitle

\section{\normalsize\bf{Introduction}}

Gravitational collapse is one of the most important problem in
classical general relativity. Usually, the formation of compact
stellar objects such as white dwarf and neutron star are preceded
by a period of collapse. The study of gravitational collapse was
started by Oppenheimer and Snyder [1]. They studied  collapse of
dust with a static Schwarzschild exterior while interior
space-time is represented by Friedman like solution. Since then
several authors have extended the above study of collapse of
which important and realistic generalizations are the following:
(i) the static exterior was studied by Misner and Sharp [2] for a
perfect fluid in the interior, (ii) using the idea of outgoing
radiation of the collapsing body by Vaidya [3], Santos and
collaborations [4-9] included the dissipation in the source by
allowing radial heat flow (while the body undergoes radiating
collapse). Ghosh and Deskar [10] have considered collapse of a
radiating star with a plane symmetric boundary (which has a close
resemblance with spherical symmetry [11]) and have concluded with
some general remarks. Also Cissoko et al [12] have studied
junction conditions between static and non-static space-times for
analyzing gravitational collapse in the presence of cosmological constant.\\

So far most of the studies have considered in a star whose interior geometry
is spherical. But in the real astrophysical situation the geometry of the interior
of a star may not be exactly spherical, rather quasi-spherical in form. Recently,
solutions for arbitrary dimensional Szekeres' model with perfect fluid (or dust)
[13] has been found for quasi-spherical or quasi-cylindrical symmetry of the space-time.
Also a detailed analysis of the gravitational collapse [14, 15] has been done for
quasi-spherical symmetry of the Szekeres' model. It has also been studied junction
conditions between quasi-spherical interior geometry of radiating star and exterior
Vaidya metric [16]. \\

In 1996, Husain [17] gave non-static spherically symmetric solutions of the Einstein
equations for a null fluid source with pressure $p$ and density $\rho$ related by $p=k\rho$.
Wang et al [18] has generalized the Vaidya solution which include most of the known solutions
to the Einstein equation such as anti-de-Sitter charged Vaidya solution. Husain solution has
been used to study the formation of a black hole with short hair [19] and can be considered
as a generalization of Vaidya solution [18]. Recently, Patil et al [20, 21] have studied the
gravitational collapse of the Husain solution in four and five dimensional space-times.\\

Recent observational data shows anisotropy in the Cosmic Microwave Background Radiation (CMBR) [22] and
the evidences from type Ia Supernovae SN 1997H redshift [23 - 25] suggest that the universe is flat and is
undergoing at present an accelerating phase preceded by a period of deceleration. But known ordinary
(baryonic) matter or radiation can not explain these observational facts, there should be a significant
amount of extraordinary non-baryonic matter (dark matter) and energy (dark energy). In last few years,
several models have been suggested to incorporate the recent observational evidences and among them, a
single component perfect fluid having exotic equation of state, known as Chaplygin gas [26, 27] is of great interest.
A further generalized form is known as modified Chaplygin gas (see eq. (16) with $B=$ constant with equation of state) (MCG) [28].
The MCG model can interpolate states between standard fluids at high density and constant -ve pressure at low density.
For $\gamma=1/3$, the MCG model describes evolution from radiation epoch to $\Lambda$CDM era at late time (where the fluid
has constant energy density and behaves as a cosmological constant).\\

Further, from phenomenological point of view MCG model is interesting and can be motivated by the brane world interpretation [29].
Moreover, this model is consistent with various classes of cosmological tests namely gravitational lensing [30], gamma-ray
bursts [31] as well as the above mentioned observations. Also for low energy density, the equation of state is that of a
polytropic gas [32] with -ve index. Therefore it is possible to have astrophysical implications of the present model with an
alternative way of restricting the parameters [33].\\

Moreover, the Chaplygin gas model produces oscillations or exponential blow up of the matter power spectrum that are
inconsistent with observation [33]. Variable Chaplygin gas model interpolates between a universe dominated by dust and
a quiessence dominated universe described by the constant equation of state. Also the model corresponds to a Born-Infeld
tachyon action [29]. Recently, the model parameters were constrained using the location of peaks of the CMB spectrum [32] and SNIa data [34].
Also the variable Chaplygin gas model is constrained using SNeIa gold data [35]. \\

The study of higher dimensional space-time model can be justified from two perspective. Firstly, inspiration comes from
superstring theory and supergravity theory where one must have higher dimensional space-time. Also the idea of brane world
scenario needs extra dimension for bulk space. Secondly, in recent past it has been shown that gravitational collapse in
higher dimension has some peculiarity at the fifth dimension. So for the present model we have investigated collapse
dynamics in higher dimensional model.\\

In this paper, we give the generalization of Husain solution in
$(n+2)$-dimensional spherically symmetric space-time for a null
fluid source with pressure $p$  and density $\rho$ are related by
the following relations (i) $p=k\rho$, (ii)
$p=k\rho-\frac{B(v)}{\rho^{\alpha}}$ (variable modified
Chaplygin) and (iii) $p=k\rho^{\alpha}$ (polytropic) [36]. For
$k=0$, (ii) reduces to variable Chaplygin model [35]. We have
also considered the particular cases namely, (i) $k=0$, $B=$
constant (generalized Chaplygin) [27], (ii) $B=$ constant
(modified Chaplygin) [28]. We have considered the interior
space-time by Szekeres' model  while for exterior
geometry we have considered generalization of Husain space-time and is presented in the appendix.\\

\section{\normalsize\bf{Basic Equations for Husain Solution in} $(n+2)$\normalsize\bf{-Dimensional Spherically Symmetric Space-Time}}

The metric ansatz in $(n+2)$-dimensional spherically symmetric space-time can be taken as

\begin{equation}
ds^{2}=-\left(1-\frac{m(v,r)}{r^{n-1}}\right)dv^{2}+2dvdr+r^{2}d\Omega_{n}^{2}
\end{equation}

where as usual $r$ is the radial co-ordinate ($0<r<\infty$), the null co-ordinate $v$
($-\infty\le v\le \infty$) stands for advanced Eddington time co-ordinate such that $r$
decreases towards the future direction along any $v=$constant trajectory, $m(v,r)$ gives
the gravitational mass inside the sphere of radius $r$ and $d\Omega_{n}^{2}$ is the line element on a unit $n$-sphere.\\

For the matter field, we have two non-interacting components namely the Vaidya null radiation
and a perfect fluid having form of the energy momentum tensor

\begin{equation}
T_{\mu\nu}=T_{\mu\nu}^{(n)}+T_{\mu\nu}^{(m)}
\end{equation}
with
\begin{equation}
T_{\mu\nu}^{(n)}=\sigma l_{\mu}l_{\nu}
\end{equation}
and
\begin{equation}
T_{\mu\nu}^{(m)}=(\rho+p)(l_{\mu}\eta_{\nu}+l_{\nu}\eta_{\mu})+pg_{\mu\nu}
\end{equation}

In the comoving co-ordinates ($v,r,\theta_{1},\theta_{2},...,\theta_{n}$), the two eigen vectors of energy-momentum
tensor namely $l_{\mu}$ and $\eta_{\nu}$ are linearly independent future pointing light-like vectors (null vectors) having components

$$
l_{a}=\delta^{0}_{a} ~,~ \eta_{a}=\frac{1}{2}\left(1-\frac{m}{r^{n-1}} \right)\delta_{a}^{0}-\delta_{a}^{1}~,
~l^{a}=\delta^{a}_{1}~,~\eta^{a}=-\delta^{a}_{0}-\frac{1}{2}\left(1-\frac{m}{r^{n-1}} \right)\delta_{1}^{a}
$$

or explicitly,

\begin{equation}
l_{\mu}=(1,0,0,...,0)~ \text{and}~  \eta_{\mu}=\left(\frac{1}{2}\left(1-\frac{m}{r^{n-1}}\right),-1,0,...,0 \right)
\end{equation}

and they satisfy the relations

\begin{equation}
l_{\lambda}l^{\lambda}=\eta_{\lambda}\eta^{\lambda}=0,~ l_{\lambda}\eta^{\lambda}=-1
\end{equation}

Here $\rho$ and $p$ are the energy density and thermodynamic
pressure while $\sigma$ is the energy density corresponding to
Vaidya null radiation. For $p=\rho=0$, the matter distribution
corresponds to null radiation flowing in the radial direction.
When $\sigma=0$, then matter reduces to degenerate type I fluid
[37] (In type I fluid, the energy-momentum tensor has a time-like
eigen vectors and 3-space-like eigen vectors). In order to
satisfy the energy conditions by the stress-energy tensor (2), the
restrictions on $\rho$, $p$ and $\sigma$ are

\begin{equation}
\text{(a)~Weak~ and~strong~energy~conditions:}~\sigma>0,~\rho\ge 0,~p\ge 0
\end{equation}
\begin{equation}
\text{(b)~Dominant~energy~conditions:}~\sigma>0,~p \ge 0,~\rho\ge p~~~~~~~~~~~
\end{equation}

The non-vanishing components of the Einstein field equations (choosing $8\pi G=1=c$)
$$
G_{\mu\nu}=T_{\mu\nu}
$$

for the metric (1) with matter field having stress-energy tensor given by (2) are

\begin{equation}
\rho=\frac{nm'}{2 r^{n}},~~p=-\frac{m''}{2 r^{n-1}}~~\text{and}~ \sigma=\frac{n\dot{m}}{2 r^{n}}
\end{equation}

where an overdot and dash stand for differentiation with respect to $v$ and $r$ respectively. In order to satisfy the energy conditions (7) and (8) we have the following restrictions on $m$ from the field equations (9):\\

(i) $m'\ge 0$, $m''\le 0$ and (ii) $\dot{m}>0$.\\

The condition (i) means that the mass function either increases with $r$ or is a constant - a natural physical requirement. The second restriction (ii) implies that matter within radius $r$ increases with time - an implosion.\\

We shall now solve the above field equations (9) for the following equation of state:\\

(a) Linear equation of state,\\

(b) Variable modified Chaplygin gas model,\\

(c) Polytropic fluid.\\

{\bf Case (a):} Linear equation of state:\\

Here the linear equation of state for the perfect fluid is taken as

\begin{equation}
p=k\rho,~~~(k,~\text{a~constant})
\end{equation}

Due to dominant energy condition $k$ should be less than unity, but for other energy conditions $k(>0)$ is unrestricted and may even be greater than unity. Now using (10) in the Einstein field equations, we have the differential equation in $m$,

\begin{equation}
m''=-\frac{knm'}{r}~,
\end{equation}

for which the explicit solution for $m$ is

\begin{equation}
m(v,r)=\left\{
\begin{array}{lll}
f(v)-\frac{g(v)}{(nk-1)r^{nk-1}}~,~~~nk\ne 1\\
\\
f(v)+g(v)log~r~,~~~nk=1
\end{array}\right.
\end{equation}

where the integration functions $f(v)$ and $g(v)$ are not totally arbitrary but are restricted by the energy conditions.\\

The expression for physical quantities are (for $nk\ne 1$)

\begin{equation}
p=k\rho=\frac{nkg(v)}{2 r^{n(k+1)}}
\end{equation}

\begin{equation}
\sigma=\frac{n}{2 r^{n}}\left[\dot{f}(v)-\frac{\dot{g}(v)}{(nk-1)r^{nk-1}} \right]
\end{equation}

From equation (13) we must have $g(v)\ge 0$ for positive pressure and density.  The dominant energy condition leads,
for $nk\ne 1$, to
$$
\dot{m}=\dot{f}(v)-\frac{\dot{g}(v)}{(nk-1)r^{nk-1}}>0
$$
Physically this means that the matter within a radius $r$ increases with time, which corresponds to an implosion. This condition
is satisfied if $\dot{f}(v)>0$ and $\left\{~ \text{either ~(a)~} \dot{g}(v)>0, k<1/n \text{~or~ (b)~} \dot{g}(v)<0, k>1/n    \right\}$.\\

Therefore the solution of the Einstein equations with matter field given by equation (2) can be written as

\begin{equation}
ds^{2}=-\left[1-\frac{f(v)}{r^{n-1}}+\frac{g(v)}{(nk-1)r^{n(k+1)-2}}\right]dv^{2}+2dvdr+r^{2}d\Omega_{n}^{2}
\end{equation}

This is termed as Husain solution in higher dimension. Similar to four dimensional Husain metric, $g(v)=0$ corresponds to usual higher dimensional Vaidya metric while for $k=\left(\frac{4}{n}-1\right)$, the solution represents Charged vaidya metric. Further, one should note that the above metric (15) is asymptotically flat for $k>1/n$ and is cosmological for $k<1/n$. \\

On the otherhand, for the solution with $k=1/n$, the energy conditions are not always satisfied for all $r$ as $\dot{m}=\dot{f}(v)+\dot{g}(v)~log~r$ becomes negative (i.e., $\sigma<0$) for very small $r$. Hence we shall not discuss this solution.\\

{\bf Case (b):}  Variable modified Chaplygin gas:\\

Here the mathematical relation (equation of state) between pressure and density is chosen as

\begin{equation}
p=k\rho-\frac{B(v)}{\rho^{\alpha}}
\end{equation}

with $k(>0),~\alpha(>0)$ being constants and $B(v)$, a function of $v$ alone. Using (16) in the field equations (9), the differential equation in $m$ becomes

\begin{equation}
m''=-\frac{nkm'}{r}+\frac{B(v)(n-1)^{\alpha+1}r^{n(\alpha+1)-1}}{(nm')^{\alpha}}
\end{equation}

which has the solution (for $\alpha\ne 1,~nk\ne 1$)

\begin{eqnarray*}
m(v,r)=f(v)-\frac{\{g(v)\}^{\frac{1}{1+\alpha}}
}{(nk-1)r^{nk-1}}~_{2}F_{1}[\frac{1-nk}{n(1+k)(1+\alpha)},
-\frac{1}{1+\alpha},1+\frac{1-nk}{n(1+k)(1+\alpha)},
\end{eqnarray*}

\begin{equation}
-\frac{B(v)(n-1)^{1+\alpha}r^{n(1+k)(1+\alpha)}}{(1+k)n^{1+\alpha}g(v)} ]
\end{equation}

Here $_{2}F_{1}$ is the usual hypergeometric function and $f(v)$ and $g(v)$ are as
before arbitrary integration functions. If one chooses $B$ to be constant then the above
equation of state (16) corresponds to modified Chaplygin gas while $k=0,~B=$constant represents
generalized Chaplygin gas and $k=0,~B=$constant, $\alpha=1$ corresponds to pure Chaplygin gas.
Further, for $B=0$, one gets back the solution (15) for linear equation of state.\\

{\bf Case (c):} Polytropic fluid:\\

The form of equation of state for such fluid is

\begin{equation}
p=k\rho^{\alpha},~(k,~\alpha~~\text{constants})
\end{equation}

So from the Einstein equations the differential equation becomes

\begin{equation}
m''=-\frac{k(nm')^{\alpha}}{(n-1)^{\alpha-1}r^{n\alpha-n+1}}
\end{equation}

As before, the solution in terms of hypergeometric function can be written as ($\alpha\ne 1$)

\begin{equation}
m(v,r)=f(v)+r~\{g(v)\}^{\frac{1}{1-\alpha}}~_{2}F_{1}[\frac{1}{n(1-\alpha)},-\frac{1}{1-\alpha},1+\frac{1}{n(1-\alpha)},\frac{k(n-1)^{1-\alpha}r^{n(1-\alpha)}}{n^{1-\alpha}g(v)} ]
\end{equation}

This solution can be obtained from the solution (18) as a particular case $k=0,~B(v)\rightarrow -k,~\alpha\rightarrow -\alpha$.\\

\section{\normalsize\bf{Study of Geodesic and Existence of Naked Singularity in Husain Space-Time}}

We shall discuss the existence of naked singularity in Husain space-time by studying radial null geodesics. In fact, we shall examine whether it is possible to have outgoing radial null geodesics which were terminated in the past at the central singularity $r=0$. The nature of the singularity (naked singularity or black hole) can be characterized by the existence of radial null geodesics emerging from the singularity. The singularity is at least locally naked if there exist such geodesics and if no such geodesics exist it is a black hole.\\

The equation for outgoing radial null geodesics can be obtained from equation (15) (for linear equation of state) by putting $ds^{2}=0$ (with $d\Omega_{n}^{2}=0$) as

\begin{equation}
2\frac{dr}{dv}=1-\frac{f(v)}{r^{n-1}}+\frac{g(v)}{(nk-1)r^{n(k+1)-2}}
\end{equation}

It can be seen easily that $r=0,~v=0$ corresponds to a singularity of the above differential equation. Suppose $X=\frac{v}{r}$ then we shall study the limiting behaviour of the function $X$ as we approach the singularity at $r=0,~v=0$ along the radial null geodesic identified above. If we denote the limiting value by $X_{0}$ then

\begin{eqnarray}
\begin{array}{c}
X_{0}\\
{}
\end{array}
\begin{array}{c}
=lim~~~~~~~~~ X \\
v\rightarrow 0~ r\rightarrow 0
\end{array}
\begin{array}{c}
=lim~~~~~~~~~ \frac{v}{r} \\
v\rightarrow 0~ r\rightarrow 0
\end{array}
\begin{array}{c}
=lim~~~~~~~~~ \frac{dv}{dr} \\
~~~v\rightarrow 0~ r\rightarrow 0
\end{array}
\begin{array}{c}
=2\left[1-\lambda X_{0}^{n-1}+\frac{\mu}{(nk-1)}~X_{0}^{n(k+1)-2} \right]^{-1} \\
{}
\end{array}
\end{eqnarray}

where the arbitrary functions $f(v)$ and $g(v)$ are chosen as

\begin{equation}
\begin{array}{c}
f(v)=\lambda v^{n-1}\\
g(v)=\mu v^{n(k+1)-2}
\end{array}
\end{equation}

Note that only the above choices of $f$ and $g$ have given
non-zero contribution to the limiting process given in equation
(23). Otherwise the tangent to the radial null geodesic will have
trivial solution.\\

If the above polynomial in $X_{0}$ has at least one positive real root then it is possible to have a radial null geodesic outgoing from the central singularity. More explicitely, $X_{0}$ is a root of the following equation in $X$:

\begin{equation}
\frac{\mu}{(nk-1)}~X^{n(k+1)-1}-\lambda X^{n}+X-2=0
\end{equation}

As exact analytic solution for $X$ is not possible for general $n$ and $k$ so we consider the following particular case, namely $n=2,~k=0$ (i.e., dust solution in four dimension). The above root equation simplifies to a quadratic equation

\begin{equation}
\lambda X^{2}+(\mu-1)X+2=0 ~~\text{i.e.}, ~ X_{0}=\frac{1-\mu \pm \sqrt{(\mu-1)^{2}-8\lambda}}{2}
\end{equation}

Hence, if $\mu<1$ and $1-\mu>2\sqrt{2\lambda}$ then $X_{0}$ is
always positive. As $f(v)$ and $g(v)$ are arbitrary function so
we can choose $\lambda$ and $\mu$ appropriately so that $X_{0}$
is positive definite. Hence, dust collapse in four dimension
always leads to naked singularity. On the other hand for any
$n(>2)$ we study the roots by numerical methods. Table I shows
the possibility of radial null geodesic in various dimensions for
different choices of parameters involved. We see that for $k=0$,
there do not exist any radial null geodesic above 5D for various
choices of other parameters, while for $k=1/3$ or $2/3$ it is
possible to
have radial null geodesics in 4D and higher dimensions.\\

On the otherhand, if we use modified Chaplygin gas instead of $p=k\rho$ then also the above root
equation (25) remains unchanged provided
$$
\begin{array}{c}
f(v)=\lambda v^{n-1}\\
g(v)=\mu^{1+\alpha}v^{(1+\alpha)\{n(1+k)-2\}}
\end{array}
$$
is chosen. Thus the nature of singularity of Husain model is
independent of the choice of the matter in the form of perfect
fluid having equation of state (i) $p=k\rho$,
(ii)$p=-\frac{B}{\rho}$ (pure Chaplygin), (iii)
$p=-\frac{B}{\rho^{\alpha}}$ (generalized Chaplygin), (iv)
$p=k\rho-\frac{B}{\rho^{\alpha}}$ (modified Chaplygin) and (v)
$p=k\rho^{\alpha}$ (polytropic). However, if we consider variable
modified Chaplygin gas having equation of state
$p=k\rho-\frac{B(v)}{\rho^{\alpha}}$ then the root equation (25)
modifies to

\begin{eqnarray*}
\frac{\mu}{(nk-1)}~X^{n(1+k)-1}~_{2}F_{1}[\frac{1-nk}{n(1+k)(1+\alpha)}, -\frac{1}{1+\alpha},1+\frac{1-nk}{n(1+k)(1+\alpha)}
\end{eqnarray*}
\begin{equation}
-\frac{B_{0}(n-1)^{1+\alpha}}{(1+k)(n\mu)^{1+\alpha}X^{n(1+k)(1+\alpha)}} ]-\lambda X^{n}+X-2=0
\end{equation}
where $B(v)=B_{0}v^{-2(1+\alpha)}$. Table II shows the nature of
the singularities in presence of variable modified Chaplygin gas.
Clearly, the nature of the roots will be different and hence
variable modified Chaplygin gas has an effect on the collapsing
process of the Husain model. For the choice $\alpha=1$, we have
peculiar behaviour in 5D and 6D, where there exist no radial null
geodesic but it is possible to have radial
null geodesic in 4D and higher dimensions. Also for $k=0$, we have similar behaviour as in table I.\\

\begin{center}
\begin{tabular}{|l|}
\hline\hline
~~~~$\lambda$~~~~~~~$\mu$~~~~~~~$k$
~~~~~~~~~~~~~~~~~~~~~~~~~~~Positive roots ($X_{0}$) \\ \hline
\\
~~~~~~~~~~~~~~~~~~~~~~~~~~~~~~~~~~~4D~~~~~~~~5D~~~~~~~~~6D~~~~~~~~~7D~~~~~~~~8D~~~~~~~~10D
\\ \hline\hline
\\
~~~0.1~~~~~0.1~~~~~2/3~~~~~~1.47479~~~1.67881~~~1.67375~~~1.62512~~~1.56933~~~1.47331
\\
\\
~~0.01~~~0.001~~~1/3~~~~~~2.05205,~~~~~$-$~~~~~~~2.12534,~~~17.1871~~~~9.99197~~~5.40875
\\
~~~~~~~~~~~~~~~~~~~~~~~~~~~~~~~91.7181~~~~~~~~~~~~~~~~4.96352
\\
~~~~~~~~~~~~~~~~~~~~~~~~~~~~~~~~~~~~~~~~~~~~~~~~~~~~~~~~36.8271
\\
\\
~~0.01~~~~0.1~~~~~1/3~~~~~~~~~~$-$~~~~~~~~~~~$-$~~~~~~~~1.246~~~~1.31917~~~1.32766~~~1.30382
\\
\\
~~~0.1~~~0.001~~~~~0~~~~~~~~2.82809~~~~~~~~~$-$~~~~~~~~~$-$~~~~~~~~$-$~~~~~~~~~~$-$~~~~~~~~~~~$-$
\\
~~~~~~~~~~~~~~~~~~~~~~~~~~~~~~~7.07191\\
\\
~~0.01~~~~0.1~~~~~~0~~~~~~~~~2.27998~~~~~~~~~$-$~~~~~~~~~$-$~~~~~~~~~$-$~~~~~~~~~$-$~~~~~~~~~~$-$
\\
~~~~~~~~~~~~~~~~~~~~~~~~~~~~~~~~87.72\\
\\
~0.001~~~~0.1~~~~~~0~~~~~~~~~2.22774~~~~~2.81434~~~~~~$-$~~~~~~~~$-$~~~~~~~~~$-$~~~~~~~~~~$-$
\\
~~~~~~~~~~~~~~~~~~~~~~~~~~~~~~~~897.772~~~~~6.50088
\\ \hline\hline
\end{tabular}
\end{center}

{\bf Table I:} Nature of the roots ($X_{0}$) of the equation (25) for various values of parameters involved.\\

\begin{center}
\begin{tabular}{|l|}
\hline\hline
~~~~$\lambda$~~~~~~~$\mu$~~~~~~~$k$~~~~~~~$\alpha$
~~~~~~~~~~~~~~~~~~~~~~~~~~~Positive roots ($X_{0}$) \\ \hline
\\
~~~~~~~~~~~~~~~~~~~~~~~~~~~~~~~~~~~~~~~~~4D~~~~~~~~5D~~~~~~~~~6D~~~~~~~~~7D~~~~~~~~8D~~~~~~~~10D
\\ \hline\hline
\\
~~0.01~~~0.001~~~~~2/3~~~~1~~~~~2.08069~~~2.11867~~~2.2056~~~3.17953~~~2.95017~~~2.36549
\\
\\
~~~0.1~~~~0.01~~~~~1/4~~~~~1~~~~~3.08023,~~~~~~$-$~~~~~~~~~~$-$~~~~~~39.0619~~~~~~~25~~~~~~~~10
\\
~~~~~~~~~~~~~~~~~~~~~~~~~~~~~~~~~~~~~6.33532
\\
\\
~~~0.1~~~~0.01~~~~~1/3~~~~~1~~~~~~3.40688~~~~~~$-$~~~~~~~~~~$-$~~~~~~17.2107~~~~9.9992~~~5.40875
\\
~~~~~~~~~~~~~~~~~~~~~~~~~~~~~~~~~~~~~~5.31946
\\
\\
~~0.01~~~~0.01~~~~~~0~~~~~~~1~~~~~2.925~~~~~~~~~$-$~~~~~~~~~~$-$~~~~~~~~~$-$~~~~~~~~~~~$-$~~~~~~~~~~$-$
\\
~~~~~~~~~~~~~~~~~~~~~~~~~~~~~~~~~~~~~~7.05688\\
\\
~~~0.1~~~~0.01~~~~~~~0~~~~~0.5~~~~2.93501~~~~~~$-$~~~~~~~~~~$-$~~~~~~~~~$-$~~~~~~~~~~$-$~~~~~~~~~~~$-$
\\
~~~~~~~~~~~~~~~~~~~~~~~~~~~~~~~~~~~~~~7.04588\\
\\
~~0.01~~~0.001~~~~~~0~~~~~0.5~~~~~2.12406~~~2.18144~~~2.40461~~~~~~$-$~~~~~~~$-$~~~~~~~~~$-$
\\
~~~~~~~~~~~~~~~~~~~~~~~~~~~~~~~~~~~~~~~97.856~~~~8.72369~~~3.38481
\\ \hline\hline
\end{tabular}
\end{center}

{\bf Table II:}  Nature of the roots ($X_{0}$) of the equation (27) in variable modified Chaplygin gas for various values of parameters.\\ \\

\section{\normalsize\bf{Conclusion}}

We have obtained exact solution in higher dimensional Husain model
for null fluid source with equation of state between $p$ and
$\rho$ are chosen as (i) $p=k\rho$, (ii)
$p=k\rho-\frac{B(v)}{\rho^{\alpha}}$  (variable modified
Chaplygin) and (iii) $p=k\rho^{\alpha}$ (polytropic). We have seen
that solutions for one equation of state can be obtained from the
other by suitable transformation of the parameters. An
interesting result has been obtained in studying the
gravitational collapse of higher dimensional Husain space-time
$-$ the final state of collapse is independent of the choice of
equation of states namely, linear, generalized Chaplygin,
modified Chaplygin, polytopic. However, the result will change
only if variable modified Chaplygin gas is considered ($B$ is a
function of $v$). From the table we see that in higher dimension,
the value of $\alpha$ becomes insignificant to determine the
positive root $X_{0}$. Finally, we have investigate the matching
conditions of
Husain model with interior Szekeres space-time in the appendix.\\

\[
{\bf APPENDIX}\]

{\normalsize\bf{Junction Conditions: Matching between
Quasi-Spherical Szekeres model and Husain Space-Time}}\\

Let $\Sigma$ be a time-like $(n+1)$-dimensional hypersurface which
divides $(n+2)$-dimensional space-time into two distinct
$(n+2)$-dimensional manifolds $V^{-}$ and $V^{+}$. The modified
version of Israel [38] by Santos [4, 5] will be used to find the
junction conditions. The geometry of the interior manifold
$V^{-}$ is given by the Szekeres space-time
\begin{equation}
ds^{2}_{-}=-dt^{2}+e^{2\alpha}dr^{2}+e^{2\beta}\sum^{n}_{i=1}dx_{i}^{2}
\end{equation}

where $\alpha$ and $\beta$ are functions of all the ($n+2$)
space-time variables. The stress-energy tensor of a non-viscous
heat conducting fluid has the expression [39]
\begin{equation}
T_{\mu\nu}=(\rho+p)u_{\mu}u_{\nu}+pg_{\mu\nu}+q_{\mu}u_{\nu}+q_{\nu}u_{\mu}
\end{equation}
where $\rho,~ p, ~q_{\mu}$ are the fluid density, isotropic
pressure and heat flow vector. We take the heat flow vector
$q_{\mu}$ to be orthogonal to the velocity vector i.e.,
$q_{\mu}u^{\mu}=0$. For comoving co-ordinate system we choose
$u^{\mu}=(1,0,0,0,...,0)$ and
$q^{\mu}=(0,q,q_{1},q_{2},...,q_{n})$ where
$q=q(t,r,x_{1},...,x_{n})$ and
$q_{i}=q_{i}(t,r,x_{1},...x_{n}),i=1,2,...,n$. Now from the
non-vanishing components of the Einstein field equation
$$
G_{\mu\nu}=T_{\mu\nu}
$$
we have the solutions of $\beta$ and $\alpha$ as [39]

\begin{equation}
e^{\beta}=R(t,r) e^{\nu(r,x_{1},x_{2},...,x_{n})}
\end{equation}
and
\begin{equation}
e^{\alpha}=\frac{R'+R\nu'}{D(t,r)}
\end{equation}

where $R$ and $D$ are functions of $t,~r$ only. Here $R$ and $D$ satisfy

\begin{equation}
2R\ddot{R}+(n-1)(\dot{R}^{2}-D^{2})+\frac{2}{n}~pR^{2}=(n-1)f(r)
\end{equation}
and
\begin{equation}
R\dot{D}=f(r)~ e^{-2\alpha}
\end{equation}

where $f(r)$ is the arbitrary function of $r$.\\

The function $\nu$ satisfies the equation

\begin{equation}
e^{-\nu}=A(r)\sum_{i=1}^{n} x_{i}^{2}+\sum_{i=1}^{n} B_{i}(r)
x_{i}+C(r)
\end{equation}

where $A, ~B_{i}, ~C$ are arbitrary functions of $r$ alone with
the restriction

\begin{equation}
\sum_{i=1}^{n} B_{i}^{2}-4AC=f(r)-1
\end{equation}

Now from Einstein's field equations we have the components of heat flux as

\begin{equation}
q=\frac{n}{R}\dot{D}~ e^{-\alpha}
\end{equation}
and
\begin{equation}
q_{i}=-\frac{\dot{D}}{D}\alpha_{x_{i}}e^{-\beta}
\end{equation}

However, from equation (33) we note that as $R$ and $D$ are
functions of $t$ and $r$ only, so $\alpha$ is independent of the
space co-ordinates $x_{i}$~'s ($i=1,2,...,n$) i.e.,
$\alpha_{x_{i}}=0$, $\forall ~i=1,2,...,n$. Hence from equation
(37) we have $q_{i}=0$ and from equation (36) we have seen that
$q=q(t,r)$ i.e., $q$ is a function of $t$ and $r$ only. Thus only
radial heat flow is possible for the above choice of the metric as
we consider.\\

However, for the co-ordinate transformation [13] $(x_{1},x_{2},...,x_{n})\longrightarrow
(\theta_{1},\theta_{2},...,\theta_{n})$:

\begin{eqnarray}\begin{array}{llll}
x_{1}=sin\theta_{n}sin\theta_{n-1}...~~ ...
sin\theta_{2}cot\frac{1}{2}\theta_{1}\\\\
x_{2}=cos\theta_{n}sin\theta_{n-1}...~~ ...
sin\theta_{2}cot\frac{1}{2}\theta_{1}\\\\
x_{3}=cos\theta_{n-1}sin\theta_{n-2}...~~ ...sin\theta_{2}cot\frac{1}{2}\theta_{1}\\\\
....~~ ...~~ ...~~ ...~~ ...~~ ...~~ ...~~ ...\\\\
x_{n-1}=cos\theta_{3}sin\theta_{2}cot\frac{1}{2}\theta_{1}\\\\
x_{n}=cos\theta_{2}cot\frac{1}{2}\theta_{1}
\end{array}
\end{eqnarray}

the metric in $V^{-}$ becomes

\begin{equation}
ds_{-}^{2}=-dt^{2}+e^{2\alpha}dr^{2}+\frac{1}{4}~e^{2\beta}cosec^{4}(\theta_{1}/2)
~d\Omega_{n}^{2}
\end{equation}

where $d\Omega_{n}^{2}$ is the metric on unit $n$-sphere.\\

For the exterior space-time $V^{+}$ we consider $(n+2)$-dimensional Husain space-time having metric

\begin{equation}
ds_{+}^{2}=-\left[1-\frac{m(v,z)}{z^{n-1}}\right]dv^{2}+2dvdz+z^{2}d\Omega_{n}^{2}
\end{equation}

where $m(v,z)$ is given in equation (12).\\

The intrinsic metric on the boundary $\Sigma$ is given by

\begin{equation}
ds_{\Sigma}^{2}=-d\tau^{2}+A^{2}(\tau)d\Omega_{n}^{2}
\end{equation}

Now Israel's junction conditions (as described
by Santos) are\\

(i)~ The continuity of the line element i.e.,
\begin{equation}
(ds^{2}_{-})_{\Sigma}=(ds^{2}_{+})_{\Sigma}=ds^{2}_{\Sigma}
\end{equation}

where $(~ )_{\Sigma}$ means the value of (~ ) on $\Sigma$.\\

(ii)~ The continuity of extrinsic curvature over $\Sigma$ gives
\begin{equation}
[K_{ij}]=K_{ij} ^{+}-K_{ij}^{-}=0~,
\end{equation}

where the explicit form of extrinsic curvature has the expression

\begin{equation}
K_{ij}^{\pm}=-n_{\sigma}^{\pm}\frac{\partial^{2}\chi^{\sigma}_{\pm}}{\partial\xi^{i}
\partial \xi^{j} }-n^{\pm}_{\sigma}\Gamma^{n}_{\mu
\nu}\frac{\partial\chi^{\mu}_{\pm}}{\partial\xi^{i}}\frac{\partial\chi^{\nu}_{\pm}}
{\partial\xi^{j}}
\end{equation}

Here $ \xi^{i}=(\tau,x_{1},x_{2},...,x_{n})$ are the intrinsic
co-ordinates to $\Sigma, ~\chi^{\sigma}_{\pm},~ \sigma
=0,1,2,...,n+1$ are the co-ordinates in $V^{\pm}$ and
$n_{\sigma}^{\pm}$ are the components of the normal vector to
$\Sigma$ in the co-ordinates $\chi^{\sigma}_{\pm}$. \\

Now for the interior space-time described by Szekeres metric the
boundary of the interior matter distribution (i.e., the surface
$\Sigma$) will be characterized by
\begin{equation}
f(r,t)=r-r_{_{\Sigma}}=0
\end{equation}
where $r_{_{\Sigma}}$ is a constant. As the vector with components
$\frac{\partial f}{\partial \chi^{\sigma}_{-} }$ is orthogonal to
$\Sigma$ so we take
$$n_{\mu}^{-}=(0,e^{\alpha},0,...,0).$$
So comparing the metric ansatzes given by equations (28) and (41)
for $dr=0$ we have from the continuity relation (42)

\begin{equation}
\frac{dt}{d\tau}=1,~~A(\tau)=\frac{1}{2}~e^{\beta}cosec^{2}(\theta_{1}/2)~~~~
\text{on}~~~r=r_{_{\Sigma}}
\end{equation}

Also the components of the extrinsic curvature for the interior
space-time are

\begin{equation}
K^{-}_{\tau\tau}=0~~~\text{and}~~~  K
^{-}_{\theta_{1}\theta_{1}}=cosec^{2}\theta_{1}K
^{-}_{\theta_{2}\theta_{2}}=....=\left[\frac{1}{4}\beta'
e^{2\beta-\alpha}cosec^{4}(\theta_{1}/2)\right]_{\Sigma}
\end{equation}

On the other hand for the exterior Husain metric described by the
equation (40) with its exterior boundary, given by
\begin{equation}
f(z,v)=z-z_{_{\Sigma}}(v) =0
\end{equation}

So the unit normal vector to $\Sigma$ is given by

\begin{equation}
n_{\mu}^{+}=\left(1-\frac{m(v,z)}{z^{n-1}}-2\frac{dz}{dv}\right)^{-1/2}\left(-
\frac{dz}{dv},1,0,...,0\right)
\end{equation}

and the non-zero components of the extrinsic curvatures are

\begin{equation}
K^{+}_{\tau\tau}=\left[-\frac{\ddot{v}}{\dot{v}}-\frac{(n-1)m\dot{v}}{2z^{n}}+\frac{\dot{v}}{2z^{n-1}}\frac{\partial{m}}{\partial{z}} \right]_{\Sigma}
\end{equation}
and
\begin{equation}
K^{+}_{\theta_{1}\theta_{1}}=cosec^{2}\theta_{1}K
^{+}_{\theta_{2}\theta_{2}}=....=\left[-z\dot{z}+z\dot{v}\left(1-\frac{m}{z^{n-1}}\right)
\right]_{\Sigma}
\end{equation}

where
$$
\dot{v}^{-2}=1-\frac{m}{z^{n-1}}-2\frac{dz}{dv}~~\text{and}~~z_{_{\Sigma}}={\tau}A
$$
results from the continuity of metric ansatz (40) on $\Sigma$ (here an overdot stands for differentiation w.r.t. $\tau$). Now the junction condition due to the continuity of the extrinsic curvature results

\begin{equation}
m(v,z)=\left(\frac{1}{2}~e^{\beta}cosec^{2}(\theta_{1}/2)\right)^{n-1} \left[1+\frac{1}{4}cosec^{4}(\theta_{1}/2)\left(e^{2\beta}\dot{\beta}^{2}-e^{2\beta-2\alpha}\beta'^{2} \right) \right]_{\Sigma}
\end{equation}

and

\begin{equation}
p=-q~e^{\alpha}+\frac{n(n-1)}{2}\left(\frac{f(r)-1}{R^{2}}\right)+\frac{2n(n-1)}{R^{2}}~sin^{4}(\theta_{1}/2)e^{-2\nu}-
\frac{n}{2}\left(\frac{\partial m}{\partial z} \right)_{\Sigma}
\end{equation}

On the boundary vanishing of the isotropic pressure does not
imply the vanishing of the heat flux. Thus for a quasi-spherical
shearing distribution of a collapsing fluid, undergoing
dissipation in the form of heat flow, the isotropic pressure on
the surface of discontinuity $\Sigma$ does not balance the
radiation. Hence in the absence of isotropic pressure there may
still be radiation on the boundary and the exterior space-time
$V^{+}$ will still be Husain space-time.\\

{\bf Acknowledgement:}\\

One of the authors (UD) is thankful to the authority of Institute
of Mathematical Sciences, Chennai, India for providing
Associateship Programme under which part of the work was carried
out. Also UD is thankful to CSIR,
Govt. of India for providing research project grant (No. 25(0153)/06/EMR-II).\\

{\bf References:}\\
\\
$[1]$  J. R. Oppenhiemer and H. Snyder, {\it Phys. Rev.} {\bf 56} 455 (1939).\\
$[2]$  C. W. Misner and D. Sharp, {\it Phys. Rev.} {\bf 136} b571
(1964).\\
$[3]$  P. C. Vaidya, {\it Proc. Indian Acad. Sci. A} {\bf 33} 264
(1951).\\
$[4]$  N. O. Santos, {\it Phys. Lett. A} {\bf 106} 296 (1984).\\
$[5]$  N. O. Santos, {\it Mon. Not. R. Astr. Soc.} {\bf 216} 403
(1985).\\
$[6]$  A. K. G. de Oliveira, N. O. Santos and C. A. Kolassis,
{\it Mon. Not. R. Astr. Soc.} {\bf 216} 1001
(1985).\\
$[7]$  A. K. G. de Oliveira, J. A. de F. Pacheco and N. O. Santos,
{\it Mon. Not. R. Astr. Soc.} {\bf 220} 405
(1986).\\
$[8]$  A. K. G. de Oliveira and N. O. Santos, {\it Astrophys. J.}
{\bf 312} 640 (1987).\\
$[9]$  A. K. G. de Oliveira, C. A. Kolassis and N. O. Santos, {\it
Mon. Not. R. Astr. Soc.} {\bf 231} 1011 (1988).\\
$[10]$  S. G. Ghosh and D. W. Deshkar, {\it Int. J. Mod. Phys. D}
{\bf 12} 317 (2003).\\
$[11]$  S. G. Ghosh and D. W. Deshkar, {\it Gravitation and
Cosmology} {\bf 6} 1 (2000).\\
$[12]$  M. Cissoko, J. Fabris, J. Gariel, G. L. Denmat and N. O.
Santos, {\it gr-qc}/9809057; S. M. C. V. Gonçalves,
{\it Class. Quantum Grav.} {\bf 18} 4517-4530 (2001).\\
$[13]$  S. Chakraborty and U. Debnath, {Int. J. Mod. Phys. D} {\bf 13} 1085 (2004); {\it gr-qc}/0304072.\\
$[14]$ P. Szekeres, {\it Phys. Rev. D} {\bf 12} 2941 (1975).\\
$[15]$ U. Debnath, S. Chakraborty and J. D. Barrow, {\it Gen. Rel. Grav.},
{\bf 36} 231 (2004); {\it gr-qc}/0305075.\\
$[16]$ U. Debnath, S. Nath and S. Chakraborty, {\it Gen. Rel. Grav.} {\bf 37} 215 (2005 ).\\
$[17]$ V. Husain, {\it Phys. Rev. D} {\bf 53} R1759 (1996).\\
$[18]$ A. Wang and Y. Wu, {\it Gen. Rel. Grav.} {\bf 31} 107 (1999).\\
$[19]$ J. D. Brown and V. Husain, {\it Int. J. Mod. Phys. D} {\bf 6} 563 (1997).\\
$[20]$ K. D. Patil and U. S. Thool, {\it Int. J. Mod. Phys. D} {\bf 14} 873 (2005).\\
$[21]$ K. D. Patil and S. S. Zade, {\it Int. J. Mod. Phys. D} {\bf 15} 1359 (2006).\\
$[22]$ M. C. Bento, O. Bertolami and A. A. Sen, {\it Phys. Lett. B}
{\bf 575} 172 (2003); {\it Phys. Rev. D} {\bf 67} 063003 (2003);
{\it Gen. Rev. Grav.} {\bf 35} 2063 (2003); D. Carturan and F.
Finelli, {\it Phys. Rev. D} {\bf 68} 103501 (2003); L. Amendola,
F. Finelli, C. Burigana and D. Carturan, {\it JCAP} {\bf 07} 005 (2003).\\
$[23]$ S. Perlmutter et al, {\it Astrophys. J.} {\bf 517} 565
(1998); A. G. Riess et al, {\it Astrophys. J.} {\bf 116} 109
(1998); P. de Bernardis et al, {\it Nature} {\bf 404} 995 (2000);
S. Hanany et al, {\it Astrophys. J.} {\bf 545} L5 (2000).\\
$[24]$ T. Padmanabhan, {\it Phys. Reports} {\bf 380} 235 (2003).\\
$[25]$ P. J. E. Peebles and B. Ratra, {\it Rev. Mod. Phys.} {\bf
75} 559 (2003).\\
$[26]$ S. Chaplygin, {Sci. Mem. Moscow Univ. Math.} {bf 21} 1
(1904).\\
$[27]$ A. Kamenshchik, U. Moschella and V. Pasquier, {\it Phys.
Lett. B} {\bf 511} 265 (2001); V. Gorini, A. Kamenshchik, U. Moschella and V. Pasquier, {\it gr-qc}/0403062.\\
$[28]$ H. B. Benaoum, {\it hep-th}/0205140; U. Debnath, A. Banerjee and S. Chakraborty, {\it Class. Quantum Grav.} {\bf 21} 5609 (2004).\\
$[29]$ M. C. Bento, O. Bertolami and A. A. Sen, {\it Phys. Dev. D}
{\bf 66} 043507 (2002); N. Bilic, G. B. Tupper and R. Viollier,
{\it Phys. Lett. B} {\bf  535} 17 (2001).\\
$[30]$ P. T. Silva and O. Bertolami, {\it Astron. Astrophys.}
{\bf 599} 829 (2003); A. Dev, D. Jain and J. S. Alcaniz, {\it Astron. Astrophys.} {\bf 417} 847 (2004).\\
$[31]$ O. Bertolami and P. T. Silva, {\it Mon. Not. R. Astron, Soc.} {\bf 365} 1149 (2006).\\
$[32]$ F. C. Santos, M. L. Bedran and V. Soares, {\it Phys. Lett.
B} {\bf 646} 215 (2007).\\
$[33]$ H. B. Sandvik, M. Tegmark, M. Zaldraiaga and I. Waga, {\it
Phys. Rev. D} {\bf 69} 123524 (2004).\\
$[34]$ M. Marker, S. Q. de Oliveira and I. Waga, {\it Phys. Lett. B} {\bf 555} 1 (2003).\\
$[35]$ Z. K. Guo and Y. Z. Zhang, {\it Phys. Lett. B} {\bf 645} 326 (2007), {\it astro-ph}/0506091; {\it astro-ph}/0509790;
G. Sethi, S. K. Singh and P. Kumar, {\it Int. J. Mod. Phys. D} {\bf 15} 1089 (2006).\\
$[36]$ K. Anguige and K.P.Tod, {\it Annals Phys.} {\bf 276} 257 (1999); U. Mukhopadhyay and S. Ray, {\it astro-ph}/0510550.\\
$[37]$ S. W. Hawking and G. F. R. Ellis, (1973) {\it The Large
Scale Structure of Space-Time}, Cambridge University Press.\\
$[38]$  W. Israel, {\it Nuovo Cimento} {\bf 44B} 1 (1966).\\
$[39]$ U. Debnath, S. Nath and S. Chakraborty, {\it Gen. Rel. Grav.} {\bf 37} 215 (2005).\\

\end{document}